\documentclass[a4paper,12pt]{article}
\usepackage{jcappub}
\usepackage{amsthm,graphicx}
\usepackage{epsfig}
\usepackage{latexsym, amssymb} 
\usepackage{amsmath}
\def\beq{\begin{equation}}
\def\eeq{\end{equation}}
\def\br{\begin{eqnarray}}
\def\er{\end{eqnarray}}
\def\benu{\begin{enumerate}}
\def\efnu{\end{enumerate}}
\def\nn{\nonumber}

\def\l{\left}
\def\r{\right}

\def\d{{\rm d}}
\def\f{\frac}

\def\ei{\eta_{\rm i}}
\def\ee{\eta_{\rm e}}

\def\vk{{\bf k}}
\def\vka{{\bf k}_{1}}
\def\vkb{{\bf k}_{2}}
\def\vkc{{\bf k}_{3}}
\def\ka{k_{1}}
\def\kb{k_{2}}
\def\kc{k_{3}}
\def\cG{{\cal G}}

\def\fnl{f_{_{\rm NL}}}
\def\Mp{\rm M_{_{\rm Pl}}}

\def\cl{C_{\ell}}
\def\psk{{ P}_{k}}
\def\psk{{P}_{\rm S}(k)}

\begin{document}
\title{Reconstruction of broad features in the primordial spectrum and inflaton potential from Planck}
 \author[a]{Dhiraj Kumar Hazra} 
 \author[a,b]{Arman Shafieloo}
 \author[c,d,e]{George F. Smoot}
 
\affiliation[a]{Asia Pacific Center for Theoretical Physics, Pohang, Gyeongbuk 790-784, Korea}
\affiliation[b]{Department of Physics, POSTECH, Pohang, Gyeongbuk 790-784, Korea}
\affiliation[c]{Lawrence Berkeley National Laboratory, Berkeley, CA 94720, USA}
\affiliation[d]{Institute for the Early Universe, Ewha Womans University Seoul, 120-750, Korea}
\affiliation[e]{Paris Centre for Cosmological Physics, Universite Paris Diderot, France}

\emailAdd{dhiraj@apctp.org, arman@apctp.org, gfsmoot@lbl.gov}

\abstract 
{With the recently published Cosmic Microwave Background data from Planck we address the optimized binning of the primordial power spectrum.
As an important modification to the usual binning of the primordial spectrum, along with the spectral amplitude of the bins, we allow the 
position of the bins also to vary. This technique enables us to address the location of the possible broad physical features in the primordial
 spectrum with relatively smaller number of bins compared to the analysis performed earlier. This approach is in fact a reconstruction method 
looking for broad features in the primordial spectrum and avoiding fitting noise in the data. Performing Markov Chain Monte Carlo analysis
 we present samples of the allowed primordial spectra with broad features consistent with Planck data. To test how realistic it is to have 
step-like features in primordial spectrum we revisit an inflationary model, proposed by A. A. Starobinsky which can address the similar features 
obtained from the binning of the spectrum. Using the publicly available code BINGO, we numerically calculate the 
local $\fnl$ for this model in equilateral and arbitrary triangular configurations of wavevectors and show that the obtained non-Gaussianity for this 
model is consistent with Planck results. In this paper we have also considered different spectral tilts at different bins to identify the cosmological 
scale that the spectral index needs to have a red tilt and it is interesting to report that spectral index cannot be well constrained up to $k \approx 0.01 {\rm Mpc^{-1}}$.
}


\maketitle

\section{Introduction}

At the current status of cosmological observation, Cosmic Microwave Background (CMB) is the most precise probe of primordial perturbations 
for a wide range of cosmological scales. WMAP~\cite{Hinshaw:2012fq} provided possible hints of large scale anomalies in the primordial power spectrum (PPS), 
that could indicate features in the primordial spectrum. However, cosmic variance limited the significance of any deviation from a feature-less 
primordial spectrum and these outliers could be simply statistical fluctuations that one can expect in any random realization of the data~\cite{Hazra:2013xva}. 
Cosmic variance also affects the recent Planck observations; however, since Planck~\cite{Planck} has detected the possible anomalies in the same scales as has been reported in WMAP, one
can argue that these features are not due to systematics in WMAP. This  motivates further investigation of the features in the power spectrum indicated by the 
Planck data. While we cannot certainly argue that these features have a real physical origin, it is still very important to study different possibilities. For
instance, it is possible to go beyond the the assumption of power-law form of the primordial spectrum and test more rigorously the possible inflationary scenarios
and other models that could produce the observed features in the angular power spectrum. Our approach in this paper is to look for the most broad features in the 
primordial perturbations by optimized binning of the primordial spectrum varying the width and positions of the bins. Allowing the bins to have different width 
and positions can be in fact analogous to reconstruction of the primordial spectrum given some minimal degrees of flexibility. 

In this approach we can be assured that we are not fitting noise and any improvement in the likelihood should be related to the broad features of the primordial spectrum.
Direct reconstruction of the primordial spectrum from the data, which works with large number of degrees of freedom compared to binned reconstruction, also hint 
towards possible features in the data~\cite{Hazra:2013xva,Paykari:2009ac,Hlozek:2011pc,Shafieloo:2003gf,reconstruction-all,Hazra:2013eva,Planck:inflation}, however without a proper 
smoothing of the primordial spectrum it is hard to distinguish noise fitting features from possible {\it real} outliers. From inflationary theories we know that a change in the 
slow roll potential in inflation can lead to features in the spectrum which can fit the data better by fitting the outliers unaddressed by the power law 
spectrum~\cite{Planck:inflation,Starobinsky:1992ts,features-all} and thereby provide a better fit. Our method of binning the spectrum also hint towards the possible 
deviations from the power law spectrum which is certainly useful in building inflationary models.

In this work we start our analysis with two bins for the primordial spectrum. We change the amplitudes of the two bins and also the position of the bins
and look for the best likelihood to the Planck data. This would allow us to find out how well we can improve the fit to the data considering limited
number of bins compared to the power-law form of PPS. We continue the procedure by introducing more number of bins for the form of PPS while for each bin
the position and amplitude are varied. We use Monte Carlo Markov Chain (MCMC) analysis to do our analysis and estimation of the position and amplitudes of the bins.
Efforts on finding optimal binning of primordial spectrum~\cite{Paykari:2009ac}~\footnote{Here the optimal binning refers to the construction of the power spectrum 
using Fisher matrix formalism with the help of the signal to noise ratio for a corresponding observation.} and analysis with small scale CMB data from Atacama Cosmology Telescope
using 20 {\it fixed} bins in the primordial spectrum~\cite{Hlozek:2011pc} have been carried out before.
We should note here about the important advantage of our approach to the usual binning of the primordial spectrum. While we are dealing with unknowns and we do 
not know if there are in fact features in the primordial spectrum or not, and if there are features where these features are located, binning of the primordial 
spectrum in an arbitrary way (like equispaced in cosmological scales) may result to diluting statistical significance. This can clearly happen if a possible 
feature is located in the middle of a bin (rather than being in two neighboring bins) so arbitrary binning may not necessarily find a possible feature in the data.
In our approach we start with two dynamical bins and while we introduce more number of bins we look at the improvements in the likelihood and where these bins 
are located. This will provide us with clear understanding of the problem and allow us to study the primordial spectrum step by step adding more complication and
without missing a broad feature. We should mention that analysis with variation of the position of the bins has been carried out before~\cite{Vazquez:2012ux} for the purpose of model selection in the context of Bayesian interpretation which is different from the line of our analysis. 

We have also considered  simple two tilt model to estimate at what scale we need to have a red tilt in the primordial spectrum. Varying the tilts in two different bins we 
address the constraint on the spectral tilt from CMB in different cosmological scales. 

In the next part of this paper we study how we can generate similar features in the PPS like two-bins model, looking at single field inflationary scenarios. 
While having sharp/discontinuous transitions between bins are in fact non-physical, we also introduce  phenomenological Tanh form of the primordial spectrum and 
we argue that in fact such form can be generated by a simple inflationary model.

Apart from the spectral amplitude and tilt, Planck data provides us with the constraints on primordial non-Gaussianity~\cite{Planck:ng} which indicates 
a Gaussian model is completely consistent with Planck. While we know inflationary models with features 
in the primordial spectrum can result to some considerable large non-Gaussianity it is important to find a way to address this issue in our analysis. We have considered a 
simple inflationary model proposed by A. A. Starobinsky~\cite{Starobinsky:1992ts} where the resultant primordial spectrum is very similar to some of our main 
phenomenological shapes of the primordial spectrum. This would also help us to estimate the extent of non-Gaussianity (specifically the bi-spectrum) we should expect from 
similar shapes of the primordial spectrum. We use the publicly available code BI-spectra and Non-Gaussianity Operator, BINGO~\cite{Hazra:2012yn} to calculate the local $\fnl$
in equilateral limit and for arbitrary triangular configurations of wavevectors for the modified Starobinsky-1992 model of inflation.

This paper is organized as follows. In the next Models and Methodology section we shall discuss the modeling of the binned power spectrum which is discontinuous 
at the bin positions and describe an equivalent smooth form  of PPS with the introduction of a Tanh step. In the same section we shall
present the Starobinsky-1992 model of inflation and model it appropriately to act as a candidate of the binned spectrum. The essentials of complete numerical computation of the 
power spectrum and bi-spectra for a canonical scalar field model is also discussed there with an aim to apply it for the modified Starobinsky model. In what follows in the results section the we shall 
discuss improvements in likelihood using the binned PPS and also the range of allowed variations in the primordial spectrum with respect to the power law. 
We shall also discuss the best fit theoretical model and the non-Gaussianities obtained. We close with a brief discussion towards the end.

%


\section{Models and methodology}~\label{sec:formalism}
In this section we shall discuss the phenomenological models we have tested, the priors on the parameters and the method of confronting them with the data. We
shall also revisit the Starobinsky model~\cite{Starobinsky:1992ts} with a modification (appeared in the public code BINGO~\cite{Hazra:2012yn}) and discuss a few essential details
on the framework of the calculation of $\fnl$.

\subsection{Binning of the spectrum : Model-A}
The power law spectrum is expressed by 2 parameters, the spectral amplitude $A_{\rm S}$ and the spectral index $n_{\rm S}$ through,
\beq
\psk=A_{\rm S}\times\l[\frac{k}{k_0}\r]^{(n_{\rm S}-1)}
\eeq
with $A_{\rm S}$ defined as amplitude at pivot scale $k_0$.
 In this analysis we have kept the position of the bins as variables and we define them as the following. 

In logarithmic scale the first bin associated to $r_{\rm b1}$ starts from a minimum wavenumber value $k_{\rm min}$ ($k_{\rm min}$ is the minimum wavenumber for a given radiative transport kernel) up to  $k_{\rm b1}$ given by:

\beq
k_{\rm b1}=\exp[\ln(k_{\rm min})+r_{\rm b1}\times[\ln(k_{\rm max})-\ln(k_{\rm min})]]~\label{eq:binposition}
\eeq

To connect $r_{\rm b1}$ appropriately with the physical scales we would like to mention that $k_{\rm min}\sim 7\times 10^{-6} {\rm Mpc^{-1}}$ 
and $k_{\rm max}\sim 0.39 {\rm Mpc^{-1}}$ for the best fit concordance model and these values do not change notably when we change the 
background parameters.

The logarithmic scale allows to dig out low-$\ell$ features more carefully since at this region the possible anomalies are most prominent.
We scan the parameter $r_{\rm b1}$ for 
values between 0 and 1 which ensures the total k-space coverage taking into account all possible combinations. For more than 2 bins the second break in the power spectrum is located at $k_{\rm b2}$ 
which depends on a new parameter $r_{\rm b2}$ following equation similar to Eq.~\ref{eq:binposition} but the $k_{\rm min}$ is replaced by the first 
bin position $k_{\rm b1}$. The different bin positions are given by, 
\begin{eqnarray}
k_{\rm b2}&=&\exp[\ln(k_{\rm b1})+r_{\rm b2}\times[\ln(k_{\rm max})-\ln(k_{\rm b1})]]\nn\\
k_{\rm b3}&=&\exp[\ln(k_{\rm b2})+r_{\rm b3}\times[\ln(k_{\rm max})-\ln(k_{\rm b2})]]\nn\\
.....&=&.....\nn\\
k_{\rm bN}&=&\exp[\ln(k_{\rm b(N-1)})+r_{\rm bN}\times[\ln(k_{\rm max})-\ln(k_{\rm b(N-1)})]]~\label{eq:morebinposition} 
\end{eqnarray}

Defining the bins in this way ensures that while scanning the parameter space of $r_{\rm b1-bN}$ (for N+1 bins) spanning from 0 to 1, we cover all different combinations of bins with different lengths. The amplitudes of bin ${\rm b1-bN}$ are given by $A_{\rm b1-bN}$ which is defined with a amplitude magnifier $m_{\rm b1}$ with respect to the amplitude of the best fit power law spectrum from Planck. We allow a broad priors for the amplitudes $\ln [m_{\rm b1}]$ to range from
-10 to 3 to allow very small and large amplitudes for different bins. For all the bins we also allow the overall tilt $n_{\rm S}$ (same for all bins) to vary. In this paper we call this model as model-A along with its number of bins (such as Model-A with 2 bins, etc). We should note that Model-A with two bins has 2 more degrees of freedom in comparison to standard power-law case and adding each extra bin requires two more degrees of freedom. 






\subsection{Variation of spectral tilts : Two bin Model-B and Model-C}

In the previous section to avoid large number of degrees of freedom, we assumed only one spectral index for all bins. In this section we consider two different kind of two bins models where each bin can have its own spectral index. 

Model-B is similar to Model-A but we assign different spectral index for each bin. In this paper we considered Model-B with only two bins and we simply call it Model-B.
Model-B has three extra degrees of freedom in comparison with standard power-law model. 

Model-C is somehow simpler than Model-B. In Model-C we have also two bins but there is no step between the two bins. These two bins can have different tilts and their 
amplitudes are equated at the transition ({\it i.e.} at $k_{\rm b1}$). Model-C looks like a broken line while its parts are still attached.  Model-C has two extra degrees of 
freedom in comparison with standard power-law model. Note that similar model was initially discussed in~\cite{Blanchard:2003du}. 

For model-B and model-C we have denoted the two spectral indices as $n_{\rm b1}$ and $n_{\rm b2}$. We have allowed red and blue tilts for both bins with $n_{\rm b}$ ranging from 0 to 2. 

These two simple phenomenological models can hint us towards any special transitional behavior in the data. As we will see later, Model-C can put a clear lower bound
on the scale that we necessarily need to have a red tilt in the primordial spectrum. 



\subsection{Tanh smoothed binned spectrum}

The sharp/discontinuous transition of power between different bins looks unrealistic and non physical. However, we can have a smooth transition between bins if we use some 
mathematical function such as hyperbolic tangent. Two avoid complications, we try only to mimic Model-A with two bins by parametrizing the primordial spectrum using Tanh, 

\beq
P_{\rm S}^{\tanh}(k) = P_{\rm S}^{\rm Plaw}(k)\times\l[1-\alpha\tanh\l[\frac{k-k_{\rm b1}}{\Delta}\r]\r]\times A_{\rm Scale}~\label{eq:tanh}
\eeq
where $P_{\rm S}^{\rm Plaw}(k)$ is the power law spectrum with amplitude fixed to its best fit value obtained in Planck analysis~\cite{Planck:cparam}. $k_{\rm b1}$ 
is exactly same as it appeared in Eq.~\ref{eq:binposition} and it denotes the position of the transition. $\alpha$ acts as a height of the step and $\Delta$ acts as a steepness/width of the step. 

To scale the base model we include another parameter $A_{\rm Scale}$ (instead of $A_s$ in the standard power-law case). In our analysis, $\alpha$ ranges from 1 to -1 so that it includes the power law model corresponding to $\alpha=0$. Moreover as we did not want the power spectrum to become negative, we ensured $1-\alpha\tanh[\frac{k-k_{\rm b1}}{\Delta}]$ remains positive 
always (with the highest possible $\alpha=1$). The value $\alpha>0$ corresponds to higher amplitude at $k<k_{\rm b1}$ and $\alpha<0$ corresponds to lower amplitude at $k<k_{\rm b1}$. We have allowed a very sharp transition through $\Delta$ so as to mimic the bin results closely. The given priors on $\ln \Delta$, ranging from -10 to 0 allow a wide range of possibilities with sharp as well as perfectly smooth crossovers. We have allowed variations of $A_{\rm Scale}$ in logarithmic scales too such that $\ln A_{\rm Scale}$ can take values from -1 to 1. 

Tanh model has three extra degrees of freedom in comparison to standard power-law and 1 extra degree of freedom in comparison to Model-A with two bins. 


We have not taken into account the effects of
tensor perturbations as they are found to be negligible to be included in temperature spectrum analysis and we focus on the scalar sector only in this work.

\subsection{Theoretical model and the essentials of non-Gaussianity}

It is always important to support a phenomenological model or a parametric form by a physical theoretical model. In the previous section we used Tanh function to mimic our two bin model and make a smooth transition between the two bins. In this section we study an inflationary scenario that can result to similar tanh feature in the primordial spectrum. This is important for two reasons. First of all it shows that such binned features are not unusual in the context of inflationary cosmology and second, we can calculate non-Gaussianity for inflationary scenarios and this can help us to estimate if such binned shaped primordial spectra are in conflict with Planck estimation of non-Gaussianity or not. 

Primordial power spectrum originates from the generation and evolution of perturbations during the epoch of inflation. We usually work with the simplest 
power law form of primordial spectrum as this is motivated from the assumption that the scalar inflaton field ($\phi$) rolls slowly down its potential $V(\phi)$ all the way during inflation. 

The background evolution of the scalar field is given by the following Klein-Gordon equation,
\beq
\ddot{\phi}+3H\dot{\phi}+\d V/\d \phi=0,~\label{eq:phi}
\eeq

where, $H$ is the Hubble parameter ($=\dot{a}/a$ with $a$ as scale factor appearing in FLRW metric) and a overdot ($\dot{}$) refers to differentiation {\it  w.r.t.} cosmic time. The slow roll
of the scalar field is usually quantified with slow roll parameters $\epsilon_i$ which are defined by,
\beq
\epsilon_{i+1}=\d\ln \epsilon_i/\d N~[i\geq 1],~\label{eq:slow-roll-param}
\eeq
where N is the number of {\it e-folds} and $\epsilon_1=-\dot{H}/H^2$. A strict slow roll would imply $\epsilon_i \ll 1$ for all {\it e-folds} during inflation. 

Quantum fluctuations of the inflaton $\phi$ induces the scalar perturbation which is represented by curvature perturbation $\cal R$, the Fourier transform of which satisfies the 
curvature perturbation equation ~\footnote{For more discussion see~\cite{Sriramkumar:2009kg} and the references therein}(Mukhanov-Sasaki equation of the following form) given by,
\beq
{\cal R}_\vk''+2\, \f{z'}{z}\, {\cal R}_\vk' + k^{2}\, {\cal R}_{\vk}=0,\label{eq:curv_pert}
\eeq
where, $z=a\dot{\phi}/H$ and ${\vk}$ is the wavevector. To get the perturbation spectrum, in practice we start integrating Eq.~\ref{eq:curv_pert} from deep inside the Hubble 
radius using Bunch-Davies initial condition to the point where modes go outside the Hubble radius (super-Hubble scales) where ${\cal R}$ freezes. From the ${\cal R}$ at super-Hubble    
scales the scalar power spectrum $\psk$ which is the two point correlation of the curvature perturbation, is calculated as $\psk=(k^3/2\pi^2)|{\cal R}_k|^2$.   

Having described the perturbation equations to derive the power spectrum, we shall now discuss the main integrals to calculate the non-Gaussianities in a model of inflation. 
Using Maldacena formalism~\cite{maldacena-2003}, the bi-spectrum which is denoted by the three point correlation function of the curvature perturbation can be described 
with the following function $G(\vka,\vkb,\vkc)$, given by,  
\begin{eqnarray}
G(\vka,\vkb,\vkc)
&\equiv & \sum_{C=1}^{7}\; G_{_{C}}(\vka,\vkb,\vkc)\nn\\
&\equiv & \Mp^2\; \sum_{C=1}^{6}\; 
\Biggl\{\l[{\cal R}_{\vka}(\ee)\, {\cal R}_{\vkb}(\ee)\,{\cal R}_{\vkc}(\ee)\r]\; 
\cG_{_{C}}(\vka,\vkb,\vkc)\nn\\ 
& &+\l[{\cal R}_{\vka}^{\ast}(\ee)\, {\cal R}_{\vkb}^{\ast}(\ee)\,{\cal R}_{\vkc}^{\ast}(\ee)\r]\;
\cG_{_{C}}^{\ast}(\vka,\vkb,\vkc)\Biggr\}
\nonumber \\ & &
+ G_{7}(\vka,\vkb,\vkc).\label{eq:G}
\end{eqnarray}
where, $\Mp^2=(8\pi G)^{-1}$ and $\cG_{_{C}}(\vka,\vkb,\vkc)$ with $C =(1,6)$ are the terms appearing in the interaction Hamiltonian and is described by the following integrals, 
\begin{eqnarray}
\cG_{1}(\vka,\vkb,\vkc)
&=&2\,i\,\int_{\ei}^{\ee} \d\eta\, a^2\, 
\epsilon_{1}^2\, \l({\cal R}_{\vka}^{\ast}\,{\cal R}_{\vkb}'^{\ast}\,
{\cal R}_{\vkc}'^{\ast}+{\rm two~permutations}\r),\label{eq:cG1}\\
\cG_{2}(\vka,\vkb,\vkc)
&=&-2\,i\;\l(\vka\cdot \vkb + {\rm two~permutations}\r)\,
\nonumber \\ & & \times
\int_{\ei}^{\ee} \d\eta\, a^2\, 
\epsilon_{1}^2\, {\cal R}_{\vka}^{\ast}\,{\cal R}_{\vkb}^{\ast}\,
{\cal R}_{\vkc}^{\ast},\label{eq:cG2}\\
\cG_{3}(\vka,\vkb,\vkc)
&=&-2\,i\,\int_{\ei}^{\ee} \d\eta \, a^2\,
\epsilon_{1}^2\, \Biggl[\l(\f{\vka\cdot\vkb}{\kb^{2}}\r)\,
{\cal R}_{\vka}^{\ast}\,{\cal R}_{\vkb}'^{\ast}\, {\cal R}_{\vkc}'^{\ast} \nonumber \\ & &
+ {\rm five~permutations}\Biggr],\label{eq:cG3}\\
\cG_{4}(\vka,\vkb,\vkc)
&=&i\,\int_{\ei}^{\ee} \d\eta\, a^2\,\epsilon_{1}\,
\epsilon_{2}'\, \l({\cal R}_{\vka}^{\ast}\,{\cal R}_{\vkb}^{\ast}\,
{\cal R}_{\vkc}'^{\ast}+{\rm two~permutations}\r),\label{eq:cG4}\\
\cG_{5}(\vka,\vkb,\vkc)
&=&\frac{i}{2}\,\int_{\ei}^{\ee} \d\eta\, 
a^2\, \epsilon_{1}^{3}\, \Biggl[\l(\f{\vka\cdot\vkb}{\kb^{2}}\r)\,
{\cal R}_{\vka}^{\ast}\,{\cal R}_{\vkb}'^{\ast}\, {\cal R}_{\vkc}'^{\ast}
\nonumber \\ & & 
+ {\rm five~permutations}\Biggr],\label{eq:cG5}\\
\cG_{6}(\vka,\vkb,\vkc) 
&=&\frac{i}{2}\,\int_{\ei}^{\ee} \d\eta\, a^2\, 
\epsilon_{1}^{3}\,
\Biggl\{\l[\f{\ka^{2}\,\l(\vkb\cdot\vkc\r)}{\kb^{2}\,\kc^{2}}\r]\, 
{\cal R}_{\vka}^{\ast}\, {\cal R}_{\vkb}'^{\ast}\, {\cal R}_{\vkc}'^{\ast}
\nonumber \\ & &
+ {\rm two~permutations}\Biggr\},\label{eq:cG6}
\end{eqnarray}
where $\eta$ is the conformal time and $\eta_i$ and $\eta_f$ denotes the beginning and end of inflation respectively. The seventh term $G_{7}(\vka,\vkb,\vkc)$ appears due to 
a field redefinition and can be expressed as,
\begin{equation}
G_{7}(\vka,\vkb,\vkc)
=\frac{\epsilon_{2}(\eta_{\rm e})}{2}\,
\l(\vert {\cal R}_{\vkb}(\eta_{\rm e})\vert^{2}\, 
\vert {\cal R}_{\vkc}(\eta_{\rm e})\vert^{2} 
+ {\rm two~permutations}\r).\label{eq:G7}
\end{equation} 
For extensive discussions on this topic see Refs.~\cite{Hazra:2012yn,maldacena-2003,ng-ncsf,martin-2012a}. The extent of the non-Gaussianity, denoted by the number $\fnl$ is 
related to $G(\vka,\vkb,\vkc)$ by the relation, 
\begin{eqnarray}
\fnl(\vka,\vkb,\vkc)
&=&-\frac{10}{3}\; (2\,\pi)^{-4}\;
\l(k_{1}\, k_{2}\,k_{3}\r)^3\;  G(\vka,\vkb,\vkc)\nn\\
& &\times\l[k_1^{3}\; {P}_{_{\rm S}}(k_2)\; {P}_{_{\rm S}}(k_3)
+{\rm two~permutations}\r]^{-1}.\label{eq:fnl-d}
\end{eqnarray}
For equilateral case $\ka=\kb=\kc=k$ and under this condition it can be shown that the first and the third term and the fifth and the sixth term differ by a constant multiplicative 
factor and can be clubbed together as ${\cal G}_{13}(k)$ and ${\cal G}_{56}(k)$ respectively.   

Equipped with the theory of calculation of power spectrum and bi-spectrum for a given model of inflation we shall now discuss the model that we have presented in this paper.



In this paper we revisit a model proposed by A. A. Starobinsky~\cite{Starobinsky:1992ts} which has been discussed before in the 
context of non-Gaussianity too~\cite{Hazra:2012yn,martin-2012a,arroja-2011-2012}. The potential in this model is linear with a break at $\phi=\phi_0$.
\begin{equation}
V(\phi) 
= \l\{\begin{array}{ll}
\displaystyle
V_0 + A_{+}\, \l(\phi-\phi_0\r)\ & {\rm for}\ \phi>\phi_0,\\
\displaystyle
V_0 + A_{-}\, \l(\phi-\phi_0\r)\ & {\rm for}\ \phi<\phi_0.
\end{array}\r.\label{eq:sm-model} 
\end{equation}
where, $A_{+/-}$ are the two slopes of the potential and the constant $V_0$ is the value of the potential at the transition. The potential has a discontinuity in its first derivative and its second derivative contains a divergent delta function.
Due to the break in the potential the field fast rolls near the transition and introduces wiggles in the primordial spectrum~\cite{Hazra:2012yn,martin-2012a,arroja-2011-2012}.
As in our case, we need to represent a power spectrum that can represent the binned/tanh spectrum closely, we needed to smooth the transition with proper function. We smooth the 
transition in the following way as appeared in the publicly available code BINGO~\cite{bingosite}:
\begin{equation}
V(\phi) 
= V_0 + \frac{1}{2}(A_++A_-)(\phi-\phi_0)+\frac{1}{2}(A_+-A_-)(\phi-\phi_0)\tanh\l[\alpha_{\rm S}(\phi-\phi_0)\r].\label{eq:sm-smooth} 
\end{equation}
Where, $\alpha_{\rm S}$ regulates the steepness of the transition which upon assuming a very large value exactly reproduces Eq.~\ref{eq:sm-model}. We have checked that tuning the $\alpha_{\rm S}$ it is possible to get power spectrum similar to the Tanh model 
as explained in Eq.~\ref{eq:tanh}. However, given this potential we also wanted to see the best fit primordial spectrum that can be obtained from the data. Using BINGO we have solved 
the background and perturbation equation without any slow-roll approximation and used an add-on (yet to be publicly available) of BINGO to incorporate the analysis in 
CAMB~\cite{cambsite,Lewis:1999bs}. We have searched for the best fit power spectrum and best fit potential parameters. Having found the best fit we have used the best fit values 
to calculate the $\fnl$ for this model in equilateral and arbitrary triangular configurations of wavevectors. We should mention that we have not included the tensor perturbations in our 
analysis as the first slow-roll parameter $\epsilon_1$ is ${\cal O}(10^{-5})$ and the tensor to scalar ratio $r\simeq16\epsilon_1$ is completely negligible to affect our analysis. Along with the 
cosmological parameters to find the best fit we have searched in the potential parameter space consisting of $V_0,A_+,A_-,\phi_0$ and $\alpha_{\rm S}$. The priors on the potential 
parameters are chosen such that they can largely cover the power spectrum generated by the priors in Tanh model.

We should mention that we have used the publicly available code CAMB~\cite{cambsite,Lewis:1999bs} to calculate the angular power spectrum and 
CosmoMC~\cite{cosmomcsite,Lewis:2002ah} with Planck likelihood~\cite{Planck:likelihood} to obtain bounds on the power spectrum and other cosmological parameters. For finding the 
best fit parameters and the likelihood we have used Powell's BOBYQA method of iterative minimization~\cite{powell}. We have compared and ensured that
the best fit obtained by MCMC and Powell's method are comparable with a difference of $\simeq 1$ which is expected with Planck likelihood (as can be 
found in Planck chains). The low-$\ell$(2-49) likelihood is calculated using the commander likelihood and the high-$\ell$ (50-2500) is calculated through CAMspec in
four different frequency channels. Along with the cosmological parameter we have also marginalized our results over total 14 nuisance parameters describing various 
foreground effects in 4 different frequencies. The priors on the cosmological parameters along with the nuisance parameters used in all the cases are same as has 
been used in the Planck analysis for the base $\Lambda$CDM model~\cite{Planck:cparam}.

\section{Results}\label{sec:results}
In this section we shall present the results of our analysis. We shall first demonstrate the improvement in fit obtained for different models, following
which we shall provide bounds on the reconstructed primordial power spectrum. Finally we shall close with the theoretical model 
that we have revisited in this paper and the non-Gaussianities generated by this model.
\subsection{Improvement in fit compared to the power law model}
\begin{figure}[!htb]
\begin{center} 
\resizebox{220pt}{170pt}{\includegraphics{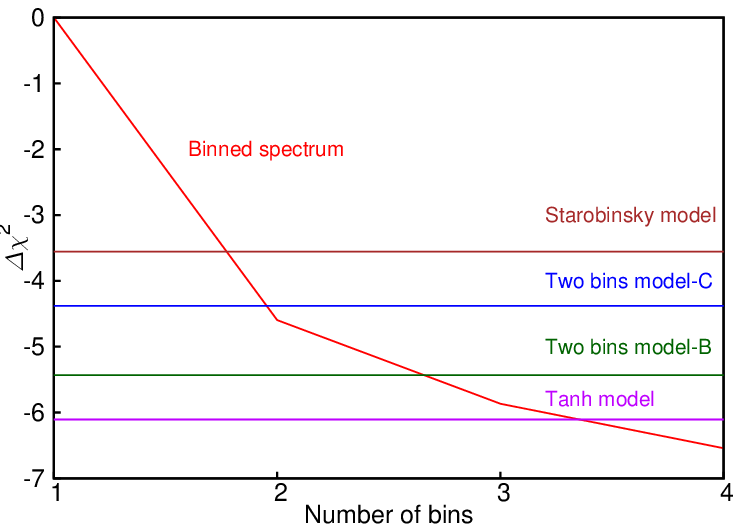}}
\end{center}
\caption{\footnotesize\label{fig:chi2} The improvement in $\chi_{\rm eff}^2$ with respect to the power law $\Lambda$CDM model as a function of number of bins is plotted. Using 
horizontal lines we have plotted the improvement obtained with different models that have been used in this analysis.}
\end{figure}

To begin with, we plot the improvement in fit obtained in all the models we have tested compared to power law primordial
spectrum. In Fig.~\ref{fig:chi2} we have plotted the difference of minimum $\chi_{\rm eff}^2$ of the model we considered and the base power law $\Lambda$CDM model. To obtain the values
we have used Powell's BOBYQA method as has been mentioned earlier. Note that with 2 bins it is possible to 
get a better fit of about $5$ over the power law model with only two extra degrees of freedom. Here we should mention that most of the improvement is coming from the low-$\ell$ commander likelihood. 
We have restricted our analysis of binned spectrum to 4 bins since considering more number of bins did not result to a significantly better likelihood to the data despite of immense computational expense. For instance, considering 4 bins introduces 8 parameters to describe the primordial spectrum (4 amplitude parameters, 3 bin position parameters and one tilt parameter) which, together with 4 background parameters and 14 nuisance parameters makes the method to find the best fit extensively time consuming. In fact we did fit the primordial spectrum with 5 bins as well but it did not improve the fit considerably beyond the 4 bin case. This is probably due to two reasons.
\begin{figure*}[!htb]
\begin{center} 
\resizebox{210pt}{160pt}{\includegraphics{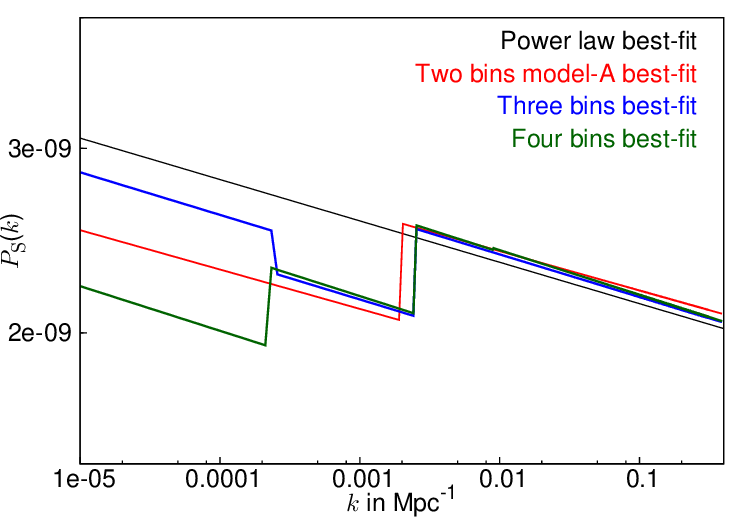}} 
\resizebox{210pt}{160pt}{\includegraphics{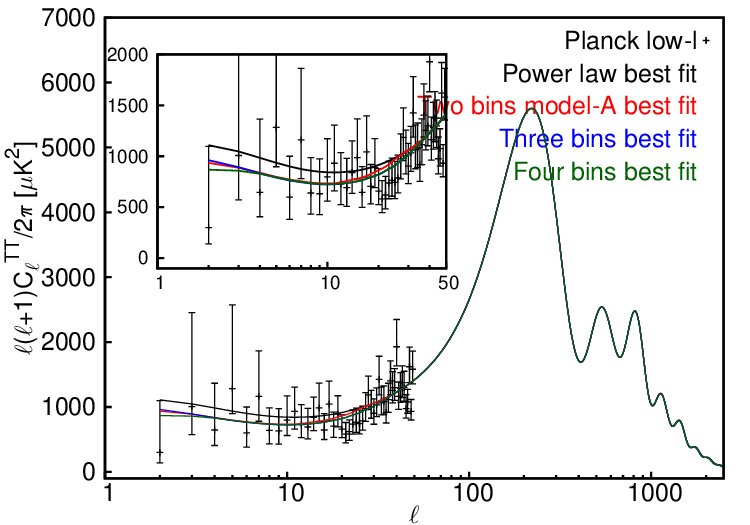}} 
\resizebox{210pt}{160pt}{\includegraphics{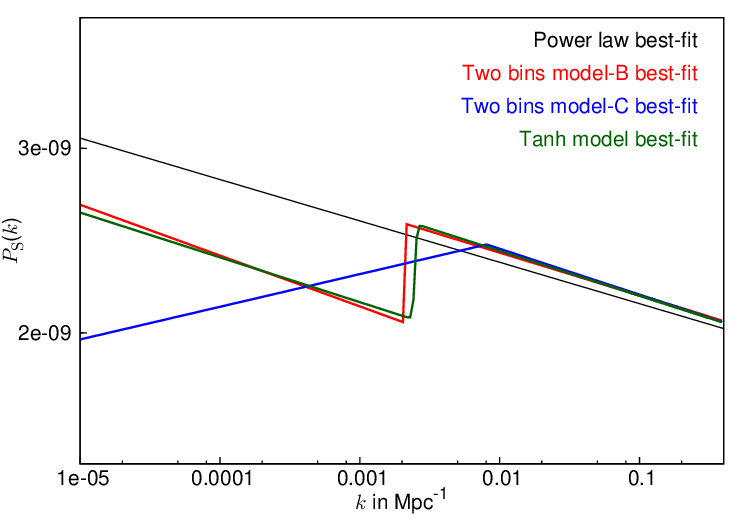}} 
\resizebox{210pt}{160pt}{\includegraphics{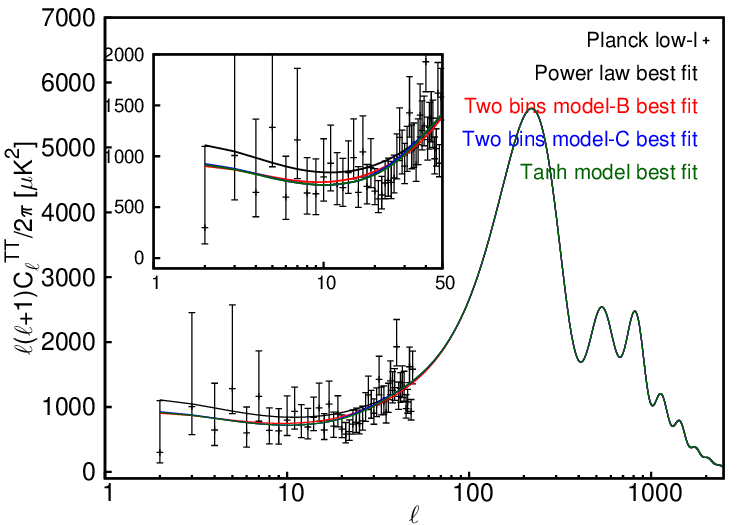}} 
\end{center}
\caption{\footnotesize\label{fig:ppscls}The best fit power spectrum and the corresponding $\cl$'s obtained from comparing the models against the Planck TT spectrum. In the upper left panel 
we present the best fit binned spectrum till 4 bins (in colors described in each of the plots) and the best fit power law spectrum (in black). In the upper right panel we have plotted the 
corresponding $\cl$'s and the Planck low-$\ell$ $\cl^{\rm TT}$ data. In the inset the Planck low-$\ell$ region is highlighted. In the lower panel we have plotted the remaining spectra and
the corresponding $\cl$'s that have been used in our analysis (except the Starobinsky model). The colors are described in the plot itself.}
\end{figure*}
Firstly due to large number of degrees of freedom the searching of the global minimum becomes inefficient. Another possibility is with 4 bins 
all the large features in the data are addressed and to get a better fit beyond that we may need sharp changes/features in the primordial spectrum which can not be addressed by the binning. We should remind here that the models with two, three and four bins have two, four and six extra degrees of freedom in comparison to standard power-law case.

In the same plot in Fig.~\ref{fig:chi2} we have located the stands of other models in the $\Delta\chi_{\rm eff}^2$ space with horizontal lines. Two bins with 2 tilts and 2 amplitudes (model-B)
fitting the data better than the two bin case as expected. Model-C which has the same degrees of freedom as two bin model-A is fitting the data slightly worse 
than model-A with two bins. Surprisingly the Tanh model is fitting the data better than the 3 bins with one less degree of freedom. 
The modified Starobinsky model has also been able to provide a better fit to the data by 3.5. We shall comment more on this towards the end of this section. 

In Fig.~\ref{fig:ppscls} we have plotted the best fit power spectrum and the corresponding $\cl^{\rm TT}$ obtained for different models in our analysis. It is interesting to see that the lack of power at low-$\ell$ suggest a step-like feature in the primordial spectrum and the position of the step is found to be at around $2\times10^{-3} {\rm Mpc}^{-1}$ in all the cases. The Tanh parametric form of the primordial spectrum also suggests a sharp transition mimicking the model with two bins. The result from Model-A with three and four bins shows that even nearly
{\it zero} power is allowed in very large scales (low-$k$ region). The Model-A with four bins shows another tiny feature near $k\simeq0.01 {\rm Mpc^{-1}}$ which
in a way indicates that there is no further scope of significant improvement in likelihood fitting the data by introducing larger number of bins in the form of PPS.
Two bins model-B does not show notably different behavior, however model-C which has 2 tilts and the spectral amplitude matches at the transition chooses a blue tilt in large scales and red tilt comparable with $n_{\rm S}=0.96$ at small scales. For two bin model-C the scale where the second 
bin starts $k\simeq0.01 {\rm Mpc^{-1}}$ implies that at least from $k_{\rm min}$ to $k\simeq0.01 {\rm Mpc^{-1}}$ a blue spectral tilt is more favored.
This is not evident from the model-B as the amplitude of the first bin compensates for the blue tilt of the model-C. So the best fit of all the models 
point to the fact that at low-$k$, till nearly $0.01 {\rm Mpc^{-1}}$ the power spectrum prefers to have a dip than a bump and at high-$k$ beyond $k> 0.01$ the power spectra 
for all the models more or less follow the power law best fit model.

\subsection{Bounds on different parameters for binning}
Having dealt with the best fit power spectrum and the improvement in fit with the featured models we shall now 
discuss the results from MCMC where we obtain the bounds and thereby present the range of allowed power spectra. 
\subsubsection{Bounds on relative amplitudes}
\begin{figure}[!htb]

\begin{center} 
\resizebox{200pt}{180pt}{\includegraphics{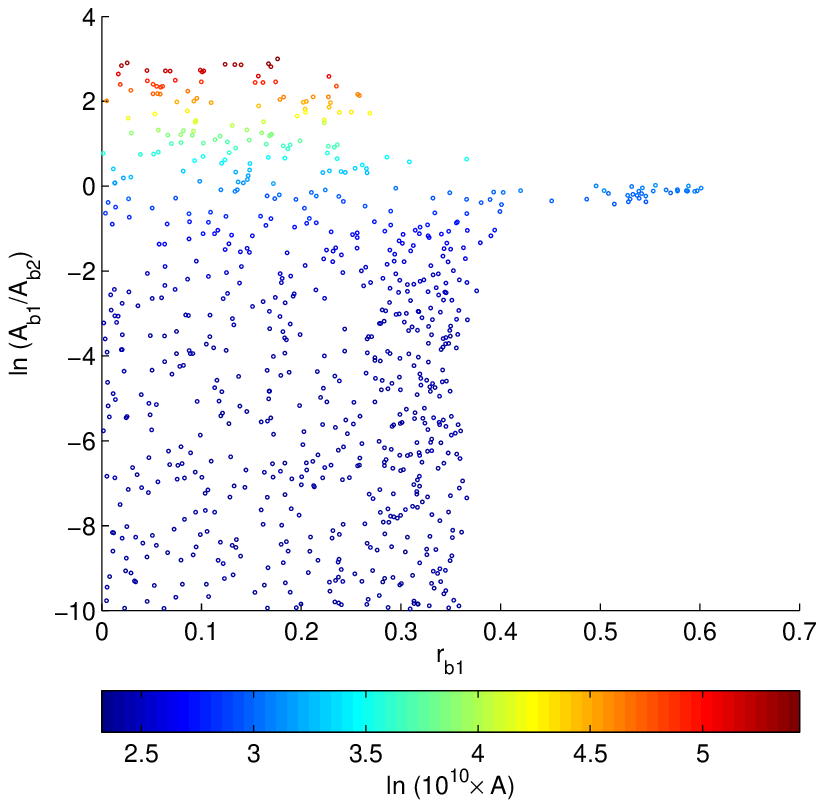}} 
\resizebox{200pt}{180pt}{\includegraphics{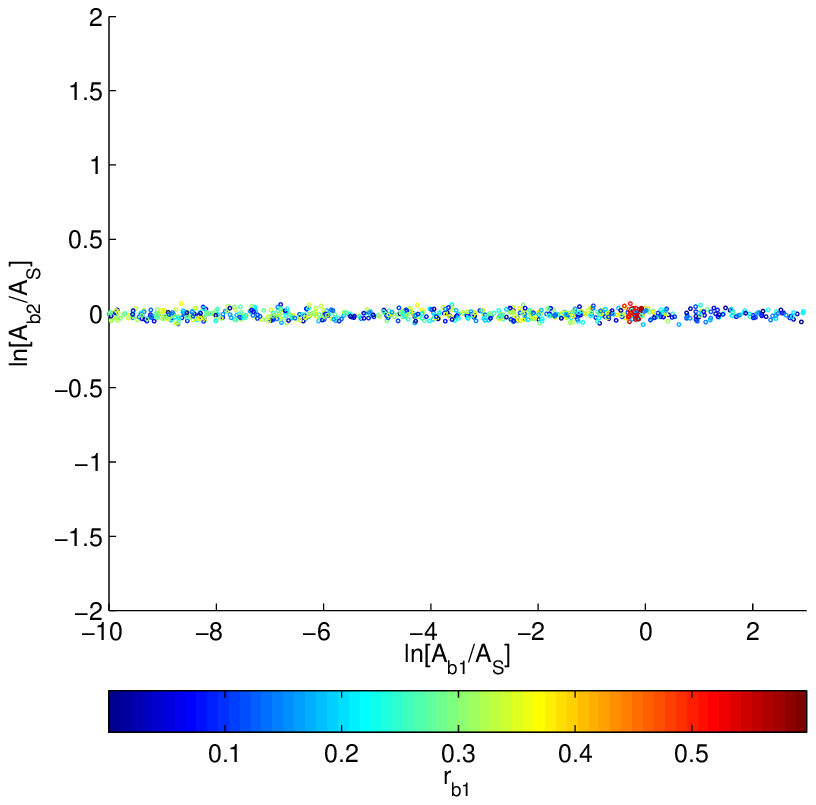}} 

\end{center}
\caption{\footnotesize\label{fig:2bin2dcntrs} For two bins, samples within 3$\sigma$ contours are plotted. In the plot appearing in the left panel we have 
plotted the logarithm of the ratio of the amplitudes of the two bins as a function of the bin position $r_{\rm b1}$ (as appears in Eq.~\ref{eq:binposition}). 
It should be noted that when the 
bin position is located near very large scales $k_{\rm min}$ (corresponding to $r_{\rm b1}\simeq 0$), the amplitude of the first bin is allowed to take very
small value compared to the second bin. The color of the sample corresponds to the quantity $\ln[10^{10}A]$ where A is the mean amplitude computed 
from the two bins. In the right panel the plotted samples correspond to the 3$\sigma$ allowed values of relative amplitudes of the first and the second bins with respect to the amplitude of the best fit power law PPS ($A_{\rm S}$). It is important to note that while the amplitude of the first bin is not constrained properly, there is a little room left to play with the amplitude of the second bin.
The color corresponds to the bin position which clearly states as the bin position nears the smallest probed scale $k_{\rm max}$ the spectrum goes back 
to power law with $\ln [A_{\rm b1}/A_{\rm S}]=\ln [A_{\rm b2}/A_{\rm S}]=0$}
\end{figure}

 In Fig.~\ref{fig:2bin2dcntrs} we have plotted the results for model-A with two bins. 
The left plot in the figure, which resembles the reconstructed band we have obtained in a previous analysis~\cite{Hazra:2013xva} shows that 
the ratio of the two amplitudes $A_{\rm b1}/A_{\rm b2}$ is not well constrained when the position of the bin is in low-$k$ region. This is due to the fact that there is nearly zero signal around $k_{\rm min}$ and $A_{\rm b1}$ can have any arbitrary value. However, as the bin position shifts to higher $k$-value the power spectrum with nearly equal amplitudes are supported by the data. It should be mentioned that there is a large degeneracy in the parameter space of 
$A_{\rm b1}/A_{\rm b2}$. For example two different spectra can have same $A_{\rm b1}/A_{\rm b2}$ (say, =1) but can be drastically different than the power 
law amplitude. To break this degeneracy we have plotted the $\ln[10^{10}A]$, where $A=[A_{\rm b1}+A_{\rm b2}]/2$, the mean amplitude. Note that, 
here too, as the bin position shifts to the higher wavenumbers, the samples become closer to the power law best fit value of $\ln[10^{10}A]=3.098$.


To look at the problem from another perspective, in the right panel of the Fig.~\ref{fig:2bin2dcntrs} we have plotted the same samples from 3$\sigma$ allowed values of the two bin amplitudes $m_{\rm b1}=A_{\rm b1}/A_{\rm S}$ and $m_{\rm b2}=A_{\rm b2}/A_{\rm S}$. Note that $A_{\rm S}$ is the best fit amplitude of the power-law PPS from Planck data.  
It is clear from the plot that the amplitude of the first bin is not really constrained within the prior range used but the amplitude of the second bin
is tightly constrained around the power law best fit value. \begin{figure}[!htb]
\begin{center} 
\resizebox{200pt}{160pt}{\includegraphics{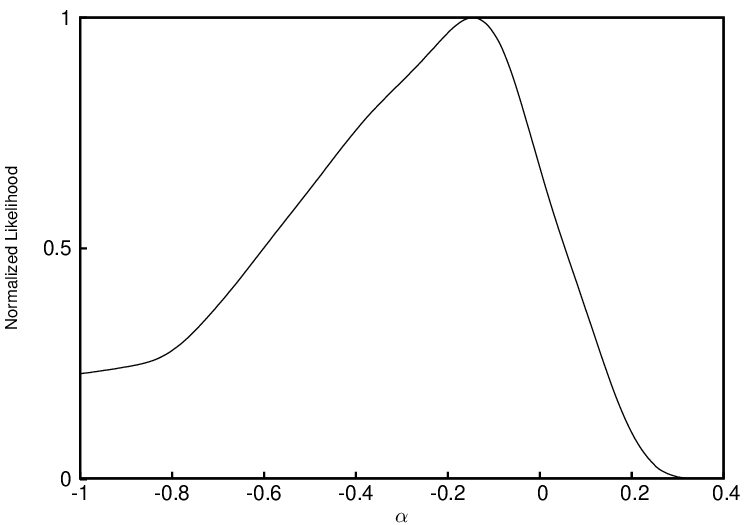}} 
\resizebox{200pt}{160pt}{\includegraphics{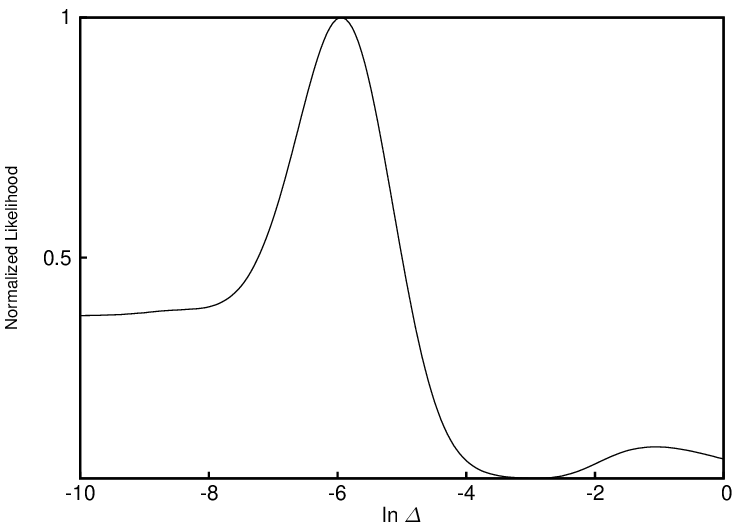}} 
\resizebox{260pt}{220pt}{\includegraphics{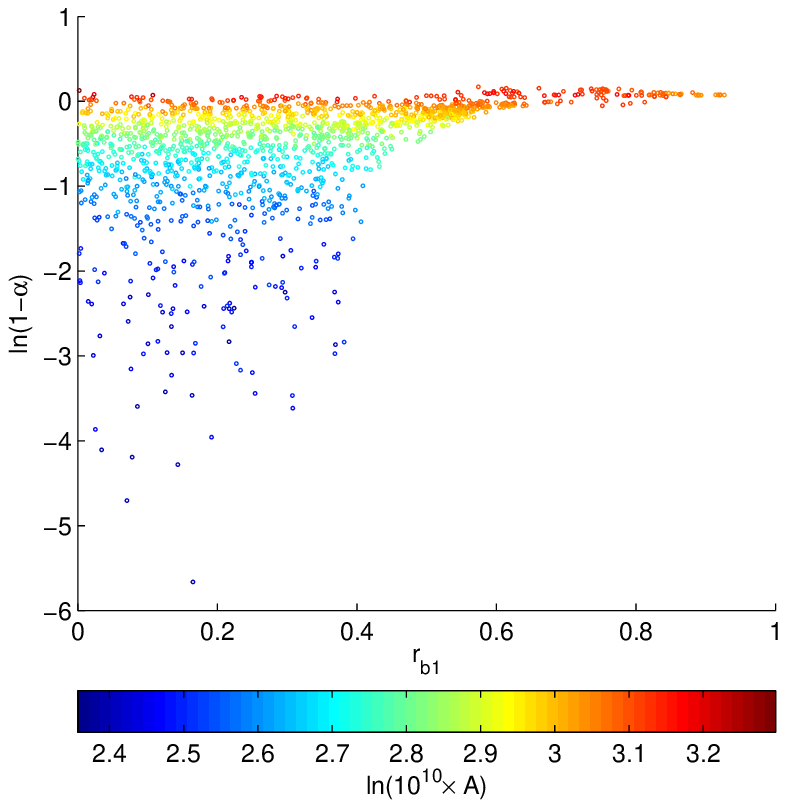}} 
\end{center}
\caption{\footnotesize\label{fig:tanhcntrs} For the Tanh model the 1-dimensional confidence contours of the strength of the Tanh step $\alpha$ and 
the steepness of the step $\Delta$(as appears in Eq.~\ref{eq:tanh}) are plotted (Upper panel). As a negative value of $\alpha$ correspond to a dip in lower-$k$ it is 
clear from the contour of $\alpha$
that a dip is much favored than a bump in large scales. At the lower panel in this figure we have plotted the samples of maximum corrections to the power law 
induced by the Tanh model within the 3$\sigma$ allowed values. Note that this plot is strikingly similar to the left plot of Fig.~\ref{fig:2bin2dcntrs}. 
The different colors again correspond to the different values of $\ln (10^{10} A)$ where A is the mean amplitude.}
\end{figure}
Though the amplitude of the first bin is unconstrained, the statement completely depends on the position of the bin. The samples in this plot has 
been colored with the values corresponding to the position of the bin. As the position of the bin approaches $k_{\rm max}$, the amplitude of the 
first bin too is constrained to a narrow region near $\ln m_{\rm b1}=0$. However, note that here the samples gather to the left of $zero$ which means a 
the data favor a dip in the amplitude of the first bin, even when the bin is positioned more than halfway ($r_{\rm b1}\simeq 0.6$) of the total $k$-range
in log scale.

In the case of Tanh model $r_{\rm b1}$ acts as a location of the step. In Fig.~\ref{fig:tanhcntrs} we have plotted the results corresponding to the Tanh model. In the upper panel of the figure we have plotted the 
marginalized 1D likelihood of the $\alpha$ and $\ln\Delta$ (in Eq.~\ref{eq:tanh}). The Tanh step is modeled in such a way (see Eq.~\ref{eq:tanh}) so that a negative value 
of $\alpha$ denotes the power in $k<k_{\rm b1}$ is lower in amplitude compared to the power in $k>k_{\rm b1}$. The plot clearly reflects that a 
dip in the larger scales is much favored than a bump. As $\ln\Delta$ is unconstrained from below it reflects that a sharp transition (say, corresponding to $\ln\Delta\simeq -10$), which 
mimics the binned power spectrum with discontinuous transition is equally probable as a smoother transition 
(say, corresponding to $\ln\Delta\simeq -6$). In the lower panel, to compare the power spectra allowed in the two bin model-A analysis 
(left plot of Fig.~\ref{fig:2bin2dcntrs}) we have plotted the $\ln(1-\alpha)$ as function of the position of the Tanh step within 3$\sigma$ confidence limit. 
As Tanh function becomes $\pm1$, for a very small value of $\Delta$, $1-\alpha$ denotes the maximum change over the power law spectrum. Note that the shape 
of both plots are in excellent agreement. Sparse samples in low $r_{\rm b1}$ region indicates that the amplitude at $k<k_{\rm b1}$ is not tightly
constrained till $r_{\rm b1}\simeq0.6$, however, notably a dip in $k<k_{\rm b1}$ is more favored. Again to break similar degeneracy explained before, we have
colored the samples with $\ln[10^{10}A]$ where $A$ correspond to the mean amplitude {\it i.e.} $A_{\rm Scale}\times A_{\rm S}$. Here too, we see that, as 
$r_{\rm b1}$ nears 1 the spectrum merges to power law best fit. Moreover we notice the constraint on the maximum allowed correction factor $1-\alpha$ 
tightens gradually with increasing $r_{\rm b1}$. Note that each of these samples corresponds to a reconstructed power spectrum which 
are in agreement with the Planck data.

\subsubsection{Bounds on spectral tilts}

\begin{figure*}[!htb]
\begin{center} 
\resizebox{220pt}{180pt}{\includegraphics{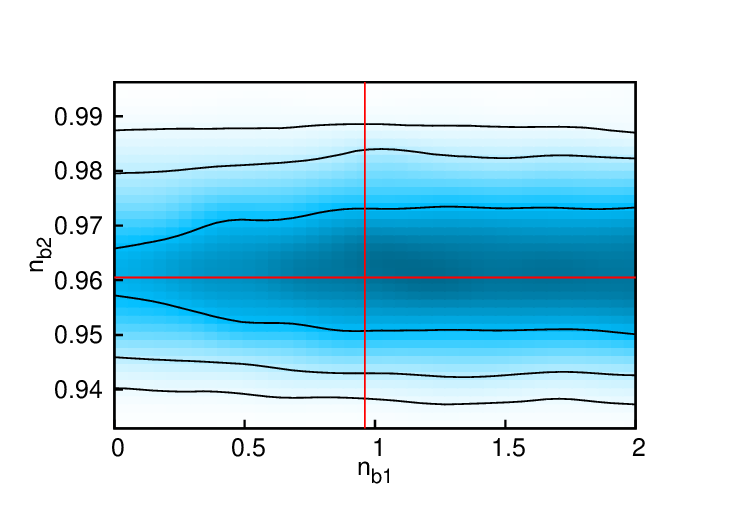}} 
\resizebox{210pt}{160pt}{\includegraphics{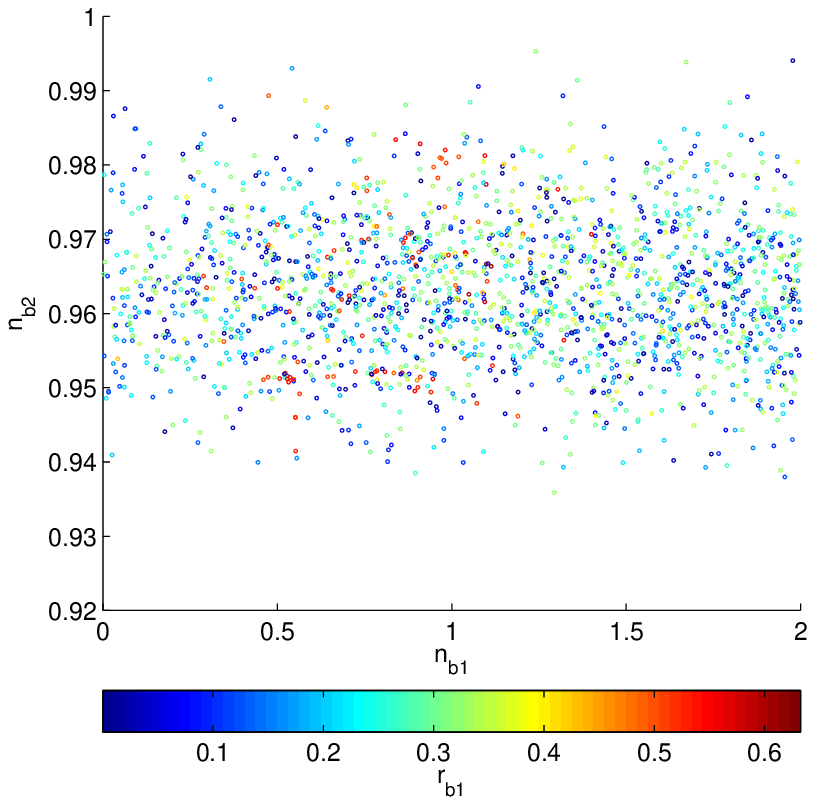}} 
\vskip -10pt
\resizebox{220pt}{180pt}{\includegraphics{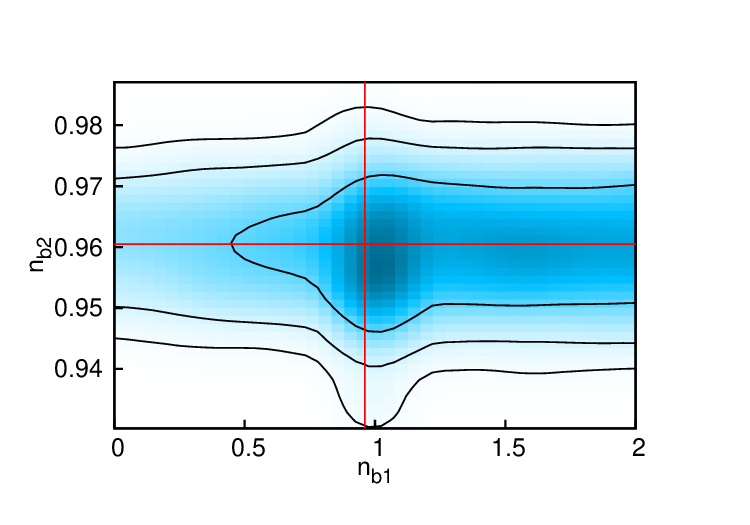}} 
\resizebox{210pt}{160pt}{\includegraphics{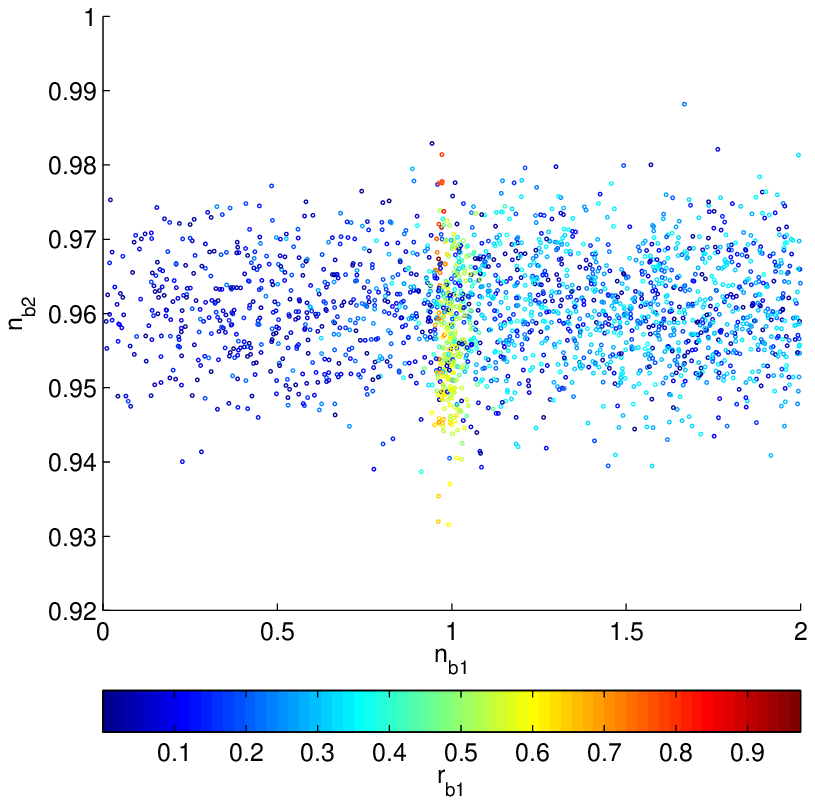}} 
\vskip -10 pt
\resizebox{220pt}{180pt}{\includegraphics{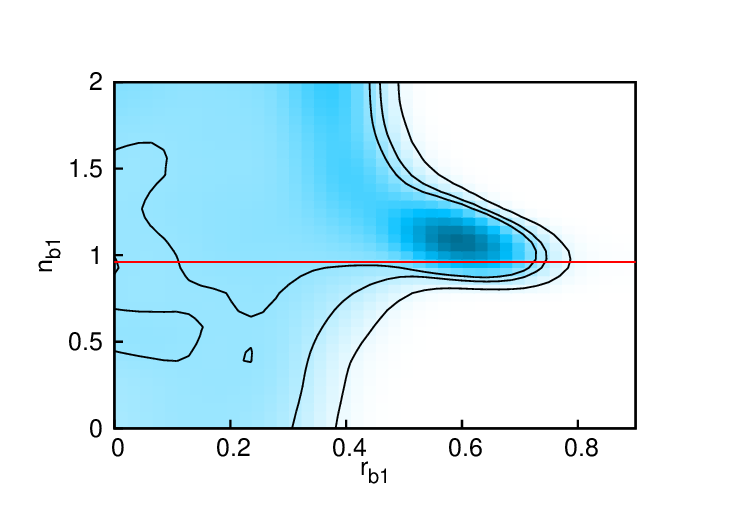}} 
\resizebox{210pt}{160pt}{\includegraphics{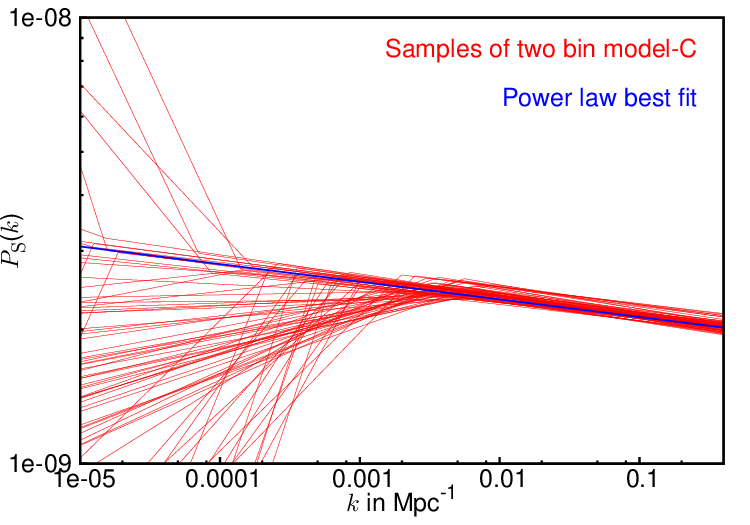}}
\end{center}
\caption{\footnotesize\label{fig:tiltcntrs} For 2 bins model-B (upper panel) and model-C (middle panel) the constraints on the two spectral tilts are 
shown. In the left panel the 3$\sigma$ confidence contours of $n_{\rm b1}$ and $n_{\rm b2}$ are plotted where we see that while the first
bin is allowed to have both red and blue tilt, the second bin must have a red tilt. For the second bin a strictly scale invariant spectrum, 
$n_{\rm b2}=1$ is not favored by the data in more than 3$\sigma$ level. In model-C as the number of amplitude decreases, due to lack of degeneracies, 
both the tilts are better constrained. In the right panel (for the plots at top and middle), for both the models the samples of $n_{\rm b1}$ and $n_{\rm b2}$ are plotted which lies 
within 3$\sigma$ CL. Samples are colored by the bin position. The left plots at the bottom corresponds to confidence contours of bin position ($r_{\rm b1}$)
and spectral index of the first bin ($n_{\rm b1}$) for two bin model-C. To its right we have plotted a few samples of the power spectrum within 2$\sigma$ 
allowed range of parameters for the same model. The red straight lines in the left plots correspond to $n_{\rm S}=0.9605$, the 
Planck best fit spectral index for power law model.}
\end{figure*}

In the analysis explained above for Model-A and Tanh case we could not study allowed range of spectral indices separately at different scales since we used only one 
spectral index for all bins.
In this regard we worked also with two bins model-B and model-C where in each bin we allow the spectral indices to vary independently. The spectral indices of the 
first and second bins are denoted by $n_{\rm b1}$ and 
$n_{\rm b2}$ respectively. In Fig.~\ref{fig:tiltcntrs} the results of the MCMC analysis are plotted. The results for model-B and model-C are plotted in 
upper and middle panels respectively. In the left panel the 3$\sigma$ contours of $n_{\rm b1}$ and $n_{\rm b2}$ are plotted (for upper and middle panel) 
which shows that the tilt of 
the first bin is unconstrained within the prior range. However, the spectral index of the second bin is constrained tightly and it rejects a completely
scale invariant spectrum (corresponding to $n_{\rm b2}=1$ at $k>k_{\rm b1}$) with more than 3$\sigma$ confidence. Model-C has one less free parameter 
than Model-B and hence constrains the spectral indices with higher CL.
A clearer picture can be provided in the right panel in the same figure for both the models. We have plotted samples from the 
left panel colored by the position of the bin. Note that the samples in model-B is more sparse than model-C. For example, when the bin position 
approaches $k_{\rm max}$ (say $r_{\rm b1}\simeq 0.6-0.7$, corresponding to ($\sim0.005-0.01{\rm Mpc^{-1}}$)) for model-B the first bin can have both 
red and blue tilt as red points are distributed in both sides of $n_{\rm b1}=1$ which implies that a worse fit imposed by a particular value of tilt 
can be compensated by the amplitude of the first bin. 

\clearpage
However, for model-C where we have only one amplitude scaling parameter we see that we loose the freedom in the first bin. Here we identify a 
clear pattern. For $r_{\rm b1}\leq 0.3$ (blue dots) the spectral tilt is allowed to be both blue and red and can not be constrained within the 
prior range. For $0.3\leq r_{\rm b1}\leq 0.5$ (cyan dots) the first bin prefers a blue till such that it creates a dip in small-$k$ region. Finally,
for $r_{\rm b1}\geq 0.5$ (green, yellow, red dots) the tilt changes gradually from a completely scale invariant to a nearly scale invariant red tilt 
of 0.96. The confidence contours of $r_{\rm b1}$ and the $n_{\rm b1}$ for model-C with two bins are plotted in the left plot at the bottom of Fig.~\ref{fig:tiltcntrs}.  
Note that till $r_{\rm b1}=0.7$ (corresponding to $k\sim 0.01 Mpc^{-1}$) the spectral tilt does not have to be red even in 1$\sigma$. Smaller scales beyond that the gradual
convergence of contours provides a clearer picture of the power of the data to constrain the spectral indicates along the cosmological scales. For an illustration,
we have also plotted a few samples of power spectrum (in red) for model-C in the bottom right panel along with the best fit power law spectrum(in blue).

\subsection{Modified Starobinsky model and the non-Gaussianity obtained}

\begin{figure*}[!t]
\begin{center} 
\resizebox{210pt}{160pt}{\includegraphics{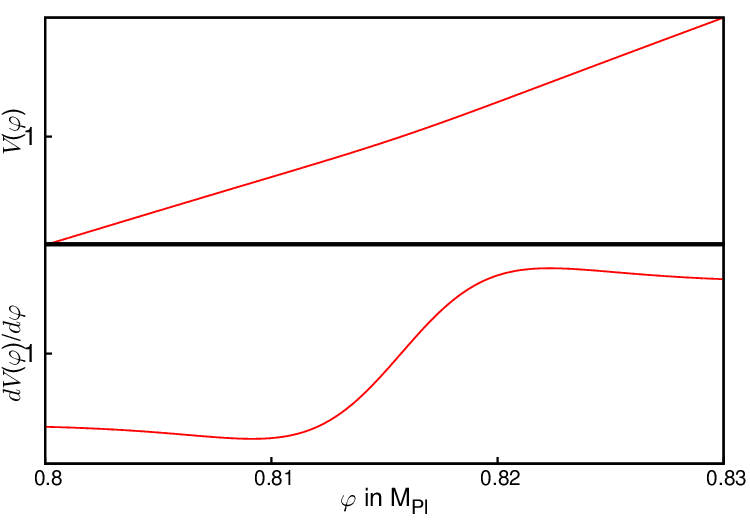}} 
\resizebox{210pt}{160pt}{\includegraphics{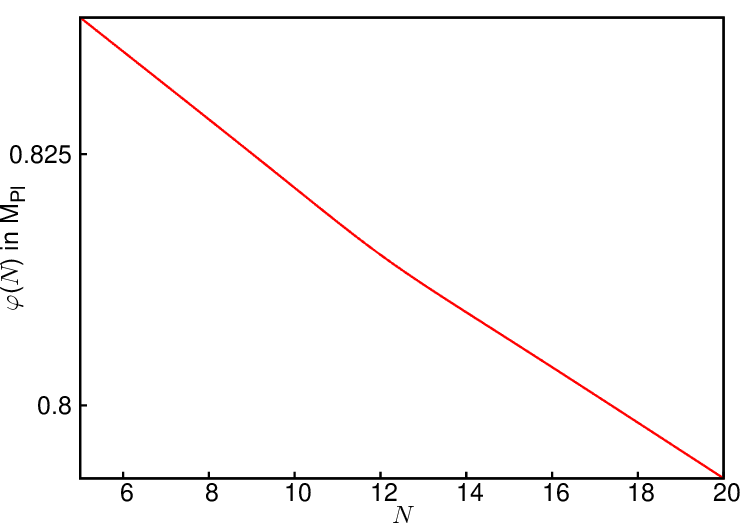}} 
\resizebox{210pt}{160pt}{\includegraphics{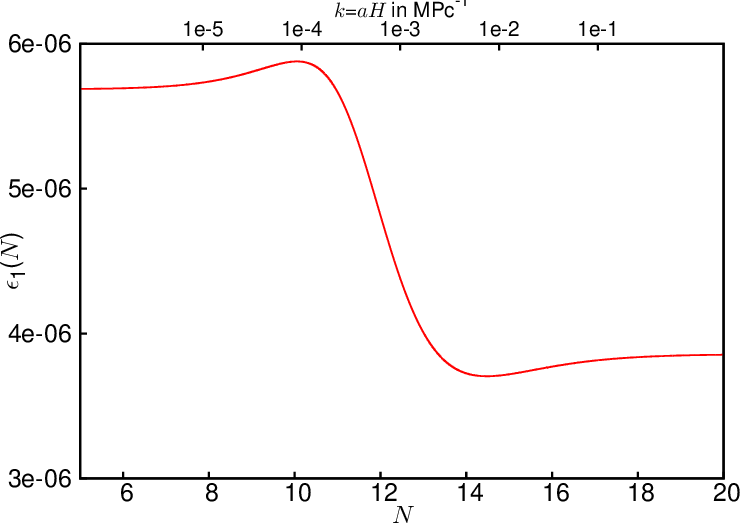}}
\resizebox{210pt}{160pt}{\includegraphics{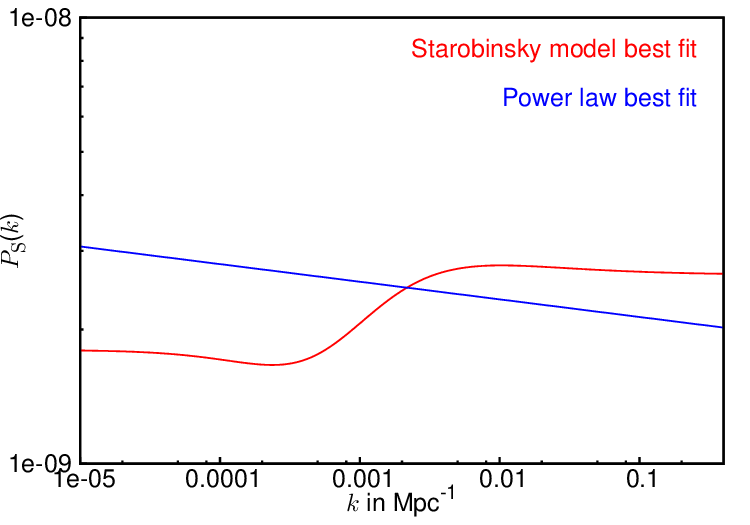}} 
\hskip -10 pt\resizebox{220pt}{160pt}{\includegraphics{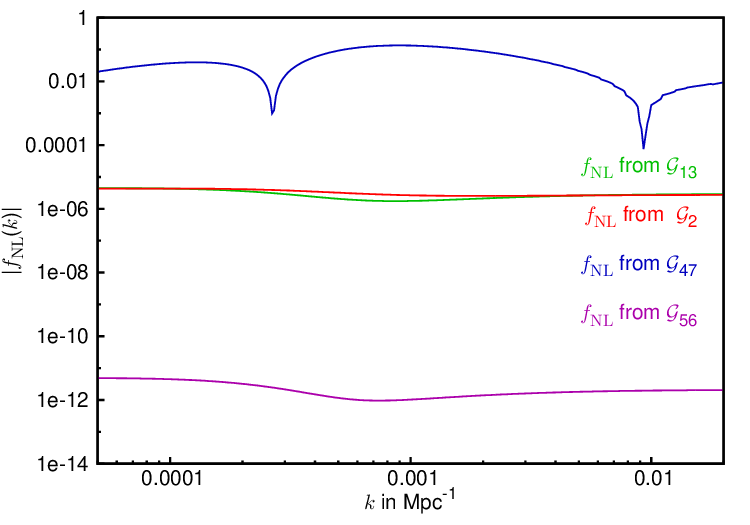}} 
\hskip -20 pt\resizebox{230pt}{185pt}{\includegraphics{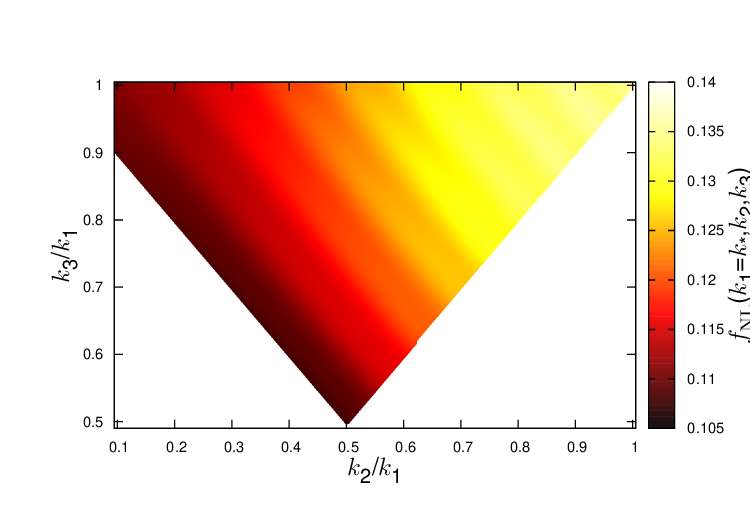}} 
\end{center}
\caption{\footnotesize\label{fig:model} The best fit results obtained upon confronting the Starobinsky model with the Planck data. In the left plot of 
the top panel we have plotted the best fit potential and its first derivative as function of the field $\phi$. Note that both the potential and 
its derivatives are normalized to 1 near at the break. In the top right panel we have plotted the behavior of the scalar field $\phi$ as a function of {\it e-folds}(N)
near the break. In left plot of the middle panel we have plotted the first slow roll parameter $\epsilon_1$ as a function of {\it e-folds}(N) near the break in the potential
where the fast roll takes place. The top axis in the same plot indicates the modes $k$ in ${\rm Mpc^{-1}}$ which leaves the Hubble radius around the corresponding {\it e-folds}
in the x-axis below. On to its right we have plotted the best fit scalar power spectrum obtained from the theoretical
model along with the best fit power law model. In the left plot on the lower panel we plot the absolute value of $\fnl(k)$ in the equilateral triangular configurations of 
wavevectors as a function of $k$ for different bi-spectrum integrals (see, Eq.~\ref{eq:cG1}-\ref{eq:cG6}) calculated from BINGO. To its right we have plotted the $\fnl(k)$ from the 
dominant contribution ${\cal G}_{47}$ for arbitrary triangular configurations of wavevectors.}
\end{figure*}

At the end of this section we shall discuss the final part of our analysis which deals with the modified Starobinsky model to explain the step-like features in the 
primordial power spectrum. As has been discussed before, we have worked with the Starobinsky-1992 model of inflation with a break in the potential 
wherein for our purpose, following previous analysis~\cite{Hazra:2012yn}, we have modeled the break as a smooth transition with Tanh function (Eq.~\ref{eq:sm-smooth}). 
In the final Fig.~\ref{fig:model} we discuss the results obtained with this  
theoretical model. In the first plot appearing in the left panel 
we have plotted the potential and its first derivative corresponding to the best fit values of the potential parameters obtained from Powell's BOBYQA method.
Note that while the change in the tilt is not clearly visible in the potential plot, its first derivative distinctly shows the transition. The best fit 
value of $\alpha_{\rm S}$ supported by the data indicates a smooth transition as shown in the plot of $\d V(\phi)/\d \phi$. Without any 
smoothing, the $\d V(\phi)/\d \phi$ from Eq.~\ref{eq:sm-model} would be discontinuous and shall have wiggles in the primordial spectrum. 
We have also plotted the behavior of the inflaton near the break in the potential in the right top plot. Here too, though a mild transition is visible, it is 
hard to estimate the amount of fast roll. As the fast roll is quantified by the slow-roll parameters, we have plotted the $\epsilon_1$ as a function of $N$  in the 
left middle plot. 
This plot clearly shows the deviation from the slow roll with a clear transition near $N\sim10-12$. In this plot the x-axis at the top 
contains the wavenumbers of the physical modes in ${\rm Mpc^{-1}}$ which leaves the Hubble radius at the corresponding $N$ provided in the x-axis at the bottom.
It is evident that these modes will be affected by the fast roll and imprint the effect in the primordial spectrum $\psk$.
To its right, we have plotted the best fit power spectrum from the model and the best fit power law spectrum. Note that modes indicated in the left plot contain 
the feature in the primordial spectrum.
We should mention that due to the dip in low-$k$ we get an 
improvement of 3.2 in the commander likelihood for $\ell=2-49$. However, from high-$\ell$ do not get any substantial improvement ($\sim0.4$) from CAMspec. 
\clearpage
We must highlight an important apparent contradiction that though a strictly scale invariant spectrum is not favored by the data with $5.5\sigma$ CL as 
reported by Planck team~\cite{Planck:cparam}, our model being scale invariant at large-$k$ is still supported by the data. In our analysis with two bin 
model-B and model-C we have shown that only from high-$k$ region, the data prefers departure from $n_{\rm S}=1$ with more than 3$\sigma$ confidence.
However a closer look at the theoretical best fit spectrum would reveal that there is indeed no discrepancy, as due to the mild bump near $0.005 {\rm Mpc^{-1}}$ 
which flattens out at small scales acts as an effective red tilt and thereby it fits high-$\ell$ the data equally well. In the left panel appearing 
at the bottom we have plotted the absolute value of local $\fnl$ as a function of $k$ for equilateral triangular configurations where, $\ka=\kb=\kc=k$. Note that, if the 
inflaton deviates from slow-roll the fourth term in the bi-spectrum integral (Eq.~\ref{eq:cG4}) becomes dominant as it contains the factor 
$\epsilon_1 \epsilon_2'$ which becomes large if a fast roll takes place. In Ref.~\cite{Hazra:2012yn} it has been demonstrated that $G_4+G_7$ 
calculated in the super-Hubble limit (with the fourth term integrated from sub-Hubble limit) is equivalent to $G_4+G_7$ evaluated till the end of 
inflation. It is clear from the plot that the maximum $\fnl^{\rm Local}$ obtained from the best fit model is ${\cal O} (0.1)$ and which is very much
consistent with the bounds on $\fnl^{\rm Local}$ obtained in Planck analysis~\cite{Planck:ng}. In the final plot of the same figure appearing in the bottom
right panel we plot $\fnl^{\rm Local}$ for different triangular configurations of wavevectors. Here we have fixed $\ka=k_{\ast}$ where $k_{\ast}$
is the particular scale where $\fnl^{\rm Local}$ reaches its maximum value in equilateral limit. Keeping $\ka$ fixed at $k_{\ast}$ we have plotted 
$\fnl^{\rm Local}$ as a function of $k_2/k_1$ and $k_3/k_1$. The color-bar represents the value of $\fnl^{\rm Local}$ which reaches its highest value at 
0.14 for the best fit model and which is again certainly consistent with Planck limits. However, we should mention that this low value of non-Gaussianities
does not throw away the possibilities of having a model with a different sharp wave features and with $\fnl^{\rm Local}\sim{\cal O} (1)$. Our main aim of this 
analysis was to locate the feature positions with the binning and present a viable theoretical model for that. 

Towards the end we would like to highlight another important fact that with Starobinsky model we find that the best fit values of cosmological 
parameters shifted considerably from the best fit values by assuming power-law form of PPS. For example the best fit $H_0$ and $\Omega_{\rm m}$ is 
found to be 70.3 and 0.277 respectively. This reflects how the derived cosmological parameters are degenerate with assumptions of the primordial spectrum~\cite{Hazra:2013eva}. 


\section{Discussion}\label{sec:discussion}

In this paper we presented an optimally binned reconstruction of primordial power spectrum using the recently 
released Planck temperature data. With up to 4 bins we addressed the locations of the possible broad features in the primordial spectrum that 
could help improving the fit to the data with respect to the power law form of the primordial power spectrum. We realized that these broad 
features in fact help us to fit better the lack of power and outliers at low multiples reported by Planck~\cite{Planck:cparam}. In this analysis we have kept the 
position of the bins as variables that enables us to locate the possible broad features in the spectrum. We report that with only two bins it is possible to get an improvement in $\chi^2_{\rm eff}$ by 5 
compared to power law best fit model (with only 2 extra degrees of freedom). Most of the improvement comes from the low-$\ell$ data and in fact 
restoring the lack of power at low-$\ell$ multipoles using the binned form of the PPS. It should be mentioned that the low-$\ell$ improvement in $\chi^2_{\rm eff}$ 
relaxes the bounds on the spectral tilt and we find that only at the high-$\ell$ (or equivalent high-$k$) the spectral tilt is bounded to be nearly scale invariant. 
With two different tilts in two bins we have identified the scales ($\sim0.01{\rm Mpc^{-1}}$) from which the data constrains the spectrum to become nearly scale 
invariant with red tilt ($n_{\rm s}<1$). We have also used a Tanh smooth form of primordial spectrum that mimic the results obtained in the two binned spectrum. Due to 
the lack of power at large scales (low multipoles) we find that a dip in the low-$k$ is in excellent agreement with the data. Basically our results suggest that till 
certain scales the amplitude and the tilt of the power spectrum is unconstrained within a considerably large limit. 
Our method being dependent of limited number of bins is unable to address high-$\ell$ anomalies ({\it if any}) as the corresponding high-$k$ space is 
more densely sampled compared to low-$k$ and one needs to have relatively large number of binning to address possible small scale features which 
is computationally expensive and not efficient due to large parameter space. While our analysis within the context of our studies shows that the power-law form of the primordial spectrum is relatively consistent to 
the Planck temperature data, to test rigorously whether the basic power law model is in perfect agreement with the data or 
not one needs to perform a proper consistency check by simulating many realizations of the data in different channels as has been 
described in~\cite{Hazra:2013xva,Hazra:2013eva} which is beyond the scope of this work. We shall try to revisit this issue in greater detail in our next work on 
reconstruction of primordial spectrum.We also reviewed Starobinsky-1992 model of inflaton with a minor modification proposed 
before and confront this model with the data. We have chosen this specific model to study since we initially expected that this model may provide us with similar features as 
the ones we obtained using our optimized binning approach. We indeed found that the best fit power spectrum obtained from this model resembles in shape to the Tanh 
model (also our Model-A with two bins). Using the publicly available code BINGO we have found that the $\fnl^{\rm Local}$ in the equilateral limit and arbitrary triangular 
configurations of wavevectors are in excellent agreement with Planck bounds on local bi-spectrum. This exercise with an actual theoretical model basically indicates that 
it is possible to have a step like form of the primordial spectrum with two different amplitudes connected by a smooth transition and generates low level of non-Gaussianities compatible with Planck results. 


\section*{Acknowledgments}
D.K.H and A.S wish to acknowledge support from the Korea Ministry of Education, Science
and Technology, Gyeongsangbuk-Do and Pohang City for Independent Junior Research Groups at 
the Asia Pacific Center for Theoretical Physics. 

\end{document}